# Soft Acoustic Curvature Sensor: Design and Development

Mohammad Sheikh Sofla, Hanita Golshanian, Vishnu Rajendran S, and Amir Ghalamzan E

*Abstract*— **This paper introduces a novel Soft Acoustic Curvature (SAC) sensor. SAC incorporates integrated audio components and features an acoustic channel within a flexible structure. A reference acoustic wave, generated by a speaker at one end of the channel, propagates and is received by a microphone at the other channel's end. Our previous study revealed that acoustic wave energy dissipation varies with acoustic channel deformation, leading us to design a novel channel capable of large deformation due to bending. We then use Machine Learning (ML) models to establish a complex mapping between channel deformations and sound modulation. Various sound frequencies and ML models were evaluated to enhance curvature detection accuracy. The sensor, constructed using soft material and 3D printing, was validated experimentally, with curvature measurement errors remaining within 3.5 m$^{-1}$ for a range of 0 to 60 m$^{-1}$ curvatures. These results demonstrate the effectiveness of the proposed method for estimating curvatures. With its flexible structure, SAC sensor holds potential for applications in soft robotics, including shape measurement for continuum manipulators, soft grippers, and wearable devices.**

## I. INTRODUCTION

IN the past decade, there has been a significant surge in the exploration of soft and elastic materials, including rubber, silicone, and hydrogels, for the advancement of robotics [1][2]. This emerging field of soft robotics harnesses the deformable properties of these materials to attain the necessary compliance for robot interactions with their environment. Unlike conventional rigid-body robots, soft robots exhibit continuous deformation across their structures, posing a distinctive challenge in real-time shape estimation to acquire the essential kinematic configuration data crucial for closed-loop control.

Traditional methods of sensing joint displacements are inapplicable to soft robots, as these robots deform uniformly along their entire length and lack discrete joints where primary displacements occur. Unlike rigid robots, which derive complete kinematic information from joint states, soft robots require shape or curvature sensing to access their kinematics. Deploying these robots beyond laboratory settings often involves obstructions or low ambient lighting conditions, rendering vision-based measurements inappropriate for estimating their shapes [3][4].

To estimate the overall shape of soft continuum robots, a kinematic model is essential, with the Piecewise Constant Curvature (PCC) approach being the most prevalent technique [5][6]. Compared to other methods, such as beam mechanics [7], the Cosserat rod model [8], and modal approaches [9], the PCC method offers advantages in terms of simplicity and suitability for real-time applications. By breaking down the robot's shape into discrete sections with constant curvatures, both modelling and control of the robot's motion are simplified. While increasing the number of sections can enhance measurement accuracy, it also increases the number of required curvature sensors and the system complexity.

Therefore, curvature sensors are vital in soft robotics for acquiring kinematic information, enabling these robots to detect and respond to changes in their shape and deformations. These sensors provide real-time feedback on the robot's current curvature, enhancing control precision and interaction with the environment. Integrating flexible curvature sensors into soft robots significantly enhances their ability to adapt to dynamic conditions and perform tasks requiring precise control of their shape and movements, particularly in situations where visual feedback is insufficient.

In the realm of curvature sensing, various materials and technologies have been explored. Based on strain measurement, different materials are examined including multilayer composites [10], conductive polymers [11], piezoceramic systems [12], and silver conductors [13]. Among these options, optical fibbers have emerged as a prevalent choice for curvature sensing. [14]. They achieve this by translating bending deformations into alterations in the intensity or phase of the optical signals transmitted along the fibre. A multitude of fibre sensing structures have been developed, such as multimode interference structures [15], long-period fibre gratings [16], and fibre Brag grating [17]. However, it is essential to note that, for many optical fibres, the sensing range is restricted to low curvature regions (curvature less than 5 m$^{-1}$) due to the inherent rigidity of the fibres [18]. Furthermore, the implementation of optical fibres often involves bulky and costly optical equipment, mainly attributed to the use of expensive laser sources and intricate

Manuscript received: May 24$^{th}$, 2024; Revised: August 5$^{th}$, 2024; Accepted: September 3$^{rd}$, 2024. This paper was recommended for publication by Editor Yong-Lae Park upon evaluation of the Associate Editor and Reviewers' comments. (Corresponding author: Amir Ghalamzan E). This work was supported in part by UKRI Research England as part of the Expanding Excellence in England (E3) Programme [Lincoln Agri-Robotics], and in part by the UK Forestry Commission [Intelligent Singulating and Labelling of Developing Trees Using Robotics (ISILDUR), #TPIF 38].

Mohammad Sheikh Sofla, Hanita Golshanian, and Vishnu Rajendran S are with the School of Agri-food Technology and Manufacturing, University of Lincoln, Lincoln, UK (e-mails: msofla@lincoln.ac.uk; hgolshanian@lincoln.ac.uk; 25451641@students.lincoln.ac.uk).

Amir Ghalamzan E is with the University of Surrey, Computer Science, Guildford, UK (email: a.ghalamzan@surrey.ac.uk).

optical interference setups.

Inertial sensing provides an alternative approach to curvature measurement. Ozel et al. [19] demonstrated the measurement of curvatures in soft-bodied bending segments using a magnet and an electronic Hall effect sensing component. MEMS-based inertial measurement units incorporate gyroscopes, accelerometers, and magnetometers, along with an Extended Kalman Filter fusion algorithm, to estimate the overall shape of a large-scale continuum manipulator [20]. However, it is important to note that the integration of rigid elements within soft bodies and susceptibility to magnetic interferences are key limitations associated with curvature estimation using inertial sensors. In [21], multi-material soft strain sensors are combined to overcome their nonlinear, time-variant behaviour, to estimate the true strain state of soft robots. However, cost-effective soft curvature sensors for providing shape feedback in soft robots are still not mature and are under exploration.

In this paper, we introduce an acoustic method for curvature estimation using a flexible, soft-sensing body. Sound is already employed in measuring a wide range of properties for fault detection [22], object recognition and classification [23][24], material finding [25], surface exploration [26], length and distance measurement [27][28], as well as force estimation and contact localization [29]-[32]. We propose a novel design for an acoustic channel composed of two flexible beams connected by thin sidewalls to estimate curvature. Acoustic waves propagate continuously through this channel, and the reduced distance between the beams due to bending significantly modulates these waves. To analyse the sound alterations, we convert the sound signal into the frequency domain using Fast Fourier Transform (FFT). We then employ a series of Machine Learning (ML) techniques to establish the relationship between the sound characteristics in the frequency domain and the curvature of the sensor body. The effectiveness of our novel curvature sensor is demonstrated through a series of test cases involving unseen curvature measurements, as presented in the results section.

To the best of our knowledge, this is the first instance of developing a Soft Acoustic Curvature (SAC) sensor. We proposed a novel Acoustic Channel design contributing to consistent good performance of the curvature sensor. The proposed sensor has several features that make it highly practical, including: (1) low cost, (2) ease of integration into soft bodies, (3) customization for various shapes and available spaces, and (4) minimal additional stiffness to the attached soft body. This curvature measurement method introduces new capabilities for soft manipulators and grippers, providing invaluable shape/curvature feedback to enhance their performance in tasks such as soft manipulation and object shape recognition.

## II. MATERIALS AND METHOD

The fundamental principle of acoustic sensing involves the modulation of sound waves as they propagate through an object. Here, we define the 'Acoustic Channel' as a hollow structure or pathway embedded within the soli/soft material. Any changes to the Acoustic Channel, such as alterations in shape or interactions with the environment, impact the transmission of sound waves, resulting in sound modulation [33]. Therefore, analysing this modulation enables the inference of the object's state. Our prior research showcased that utilizing a deformable acoustic channel, which responds to changes in the membrane's deformation, can significantly improve sensing accuracy [30]. In particular, the study utilised the deformation of a circular channel induced by applying concentrated normal force to detect the position and magnitude of the external force.

The objective of our study here is to detect curvature by utilizing a soft acoustic channel. We aim to identify the acoustic channel that has the greatest impact on the modulation of the traveling acoustic wave through the channel under different curvatures. Our scientific question is (1) "Can we use Acoustic-based soft sensor for curvature estimation?" and (2) "What are the specifications, i.e. design of the channels and sensor, of such sensor?". To answer these questions, we conducted a series of tests to investigate the effects of channel's shape on sound modulation.

### A. Acoustic attenuation analysis

Various parameters affect sound signal attenuation in an acoustic channel. At higher frequencies, the amplitude of sound signals experiences greater reduction after passing through the channel due to several factors, including higher energy loss and increased interaction with the medium's molecules [34][35]. The distance sound travels also plays a crucial role, with longer distances causing more attenuation of the signal strength [36]. Additionally, the material composition of the channel impacts sound wave absorption and reflection. Softer materials generally absorb more sound, whereas stiffer materials reflect sound waves with less attenuation [37]. Different channel shapes, even with the same cross-sectional area, can lead to varying levels of sound attenuation due to differences in boundary layer interactions and surface area contact [38]. Nonetheless, this paper is focused on a design of acoustic channel where changes in the channel's curvature significantly alter the acoustic signal attenuation for SAC sensor. After analysis and experimental investigations, we found that variation of the channel's cross-sectional area can majorly alter the acoustic signal attenuation.

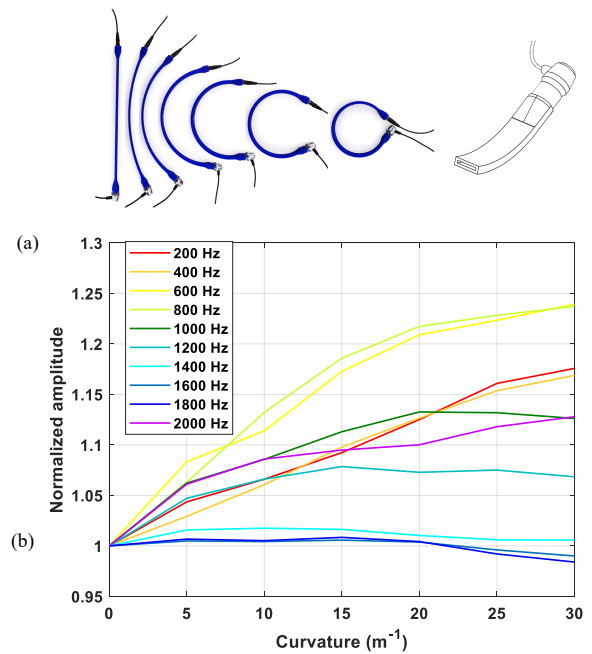

**Fig. 1.** (a) The rectangular $AC_R$ with various curvatures, and the (b) changes in amplitude of different frequencies with variations in channel curvature.

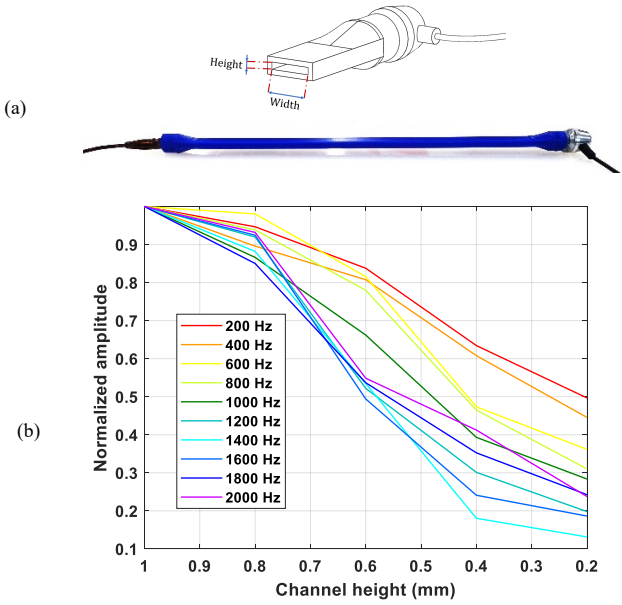

(a)

(b)

**Fig. 2.** (a) The rectangular $AC_R$ and the (b) changes in amplitude of different frequencies with variations in channel height.

To investigate the acoustic attenuation with changes in the channel's curvature and cross-sectional area, we employed PLA 3-D printing and fabricated two types of test acoustic channels embedded in rigid material ($AC_R$): (1) $AC_R$ with a fixed rectangular shape and identical area (width = 8 mm, height = 1 mm) with various curvatures ranging from 0 to 30 $m^{-1}$ (Fig. 1a), and (2) $AC_R$ with a rectangular shape and various cross-sectional areas, with a fixed curvature of 0 $m^{-1}$, an 8 mm width, and heights varying from 1 to 0.2 mm (Fig. 2a). We conducted tests using a speaker and microphone across frequencies ranging from 200 Hz to 2000 Hz. The measured amplitudes after FFT processing are plotted versus curvature and cross-sectional area in Figs. 1b and 2b, respectively. To ensure a fair comparison, we normalized the recorded amplitudes by dividing amplitude by the corresponding amplitude recorded at curvature of 0 $m^{-1}$ and a 1 mm channel height.

Our findings (shown in Fig. 1a) indicate that while increasing the channel's curvature can lead to an increase in normalized sound amplitudes at certain frequencies, it can also result in reduced or constant normalized sound amplitudes at other frequencies. This suggests that changing only the curvature of $AC_R$ does not consistently impact the sound modulation in a manner useful for reliable curvature sensing. On the other hand, our results in Fig. 2 reveal that changes in the cross-sectional area of the acoustic channel can effectively modulate sound amplitudes. Specifically, the normalized sound amplitudes at all frequencies reduce as the height of the channels decreases, with some frequencies experiencing amplitude reductions of up to 87%. The greater the modulation of sound amplitude, the richer the features available in the sound received by the microphone for a regression method to estimate curvature. Our goal is to select a sensor that provides the most informative readings for curvature sensing. Therefore, we leverage both bending and changes in the cross-sectional area of the channel. We require a channel that undergoes significant changes in cross-sectional area proportional to the bending and resulting curvature. Thus, the objective of the proposed channel design is to ensure that changes in the curvature of the sensor body also affect the area of the acoustic channel.

### B. Soft acoustic channel design

Our results from the previous section suggest a design for acoustic channels embedded in soft material ($AC_S$) that offers low resistance to bending and significantly reduces the cross-sectional area. We propose an $AC_S$ structure incorporating two thin flexible beams clamped at the ends and connected by curved sidewalls, as illustrated in Fig. 3. The lower-beam comes into contact with the surface whose curvature needs to be measured or can be integrated with the body of a soft robot.

We employ Cosserat rod theory [39][40] to analyse the deformation of the beams. In this model, the shape of the beams can be described by their centreline curve Cartesian position in space $\boldsymbol{P}(s) \in \mathbb{R}^3$, and the rotation matrix of their material orientation $\boldsymbol{R}(s) \in$ SO(3), as functions of the arc-length of the beam $s \in [0\ L]$ ($L$ is the length of undeformed beam). This model, in static form, is described by the following ordinary differential equations:

$$
\begin{aligned}
\dot{\boldsymbol{P}} &= \boldsymbol{R}(\boldsymbol{K}_{se}{}^{-1}\boldsymbol{R}^{\mathsf{T}}\boldsymbol{n} + \boldsymbol{v}^*)\,, \\
\dot{\boldsymbol{R}} &= \boldsymbol{R}(\boldsymbol{K}_{bt}{}^{-1}\boldsymbol{R}^{\mathsf{T}}\boldsymbol{m})^{\wedge}, \\
\dot{\boldsymbol{n}} &= -\boldsymbol{\rho}\,, \\
\dot{\boldsymbol{m}} &= -\dot{\boldsymbol{P}} \times \boldsymbol{n},
\end{aligned}
\tag{1}
$$

where the dot denotes a derivative with respect to $s$, and $(.)^{\wedge}$ represents the mapping from $\mathbb{R}^3$ to $\mathfrak{so}(3)$ [41]. $\boldsymbol{n}(s)$ and $\boldsymbol{m}(s)$ are the internal force and moment vectors in the global coordinate frame, respectively. $\boldsymbol{K}_{se}$ is the stiffness matrix for shear and extension, and $\boldsymbol{K}_{bt}$ represent the stiffness matrix for bending and twisting [39]. $\boldsymbol{\rho}$ corresponds to the force variations along the beams. In the proposed AC$_S$ structure, the ends of the beams are clamped together. Therefore, the following boundary condition equations for the end pose of the beams can be obtained:

$$
\begin{aligned}
\boldsymbol{P}_u(L) &= \boldsymbol{P}_l(L) \\
[\log{(R_u(L)R_l(L))}]^{\vee} &= \boldsymbol{0}
\end{aligned}
\tag{2}
$$

where the suffix $u$ stands for the upper-beam, and $l$ for the lower-beam. $\log{()}$ is the matrix natural logarithm which maps SO(3) to $\mathfrak{so}(3)$, and $(.)^{\vee}$ represents the mapping from $\mathfrak{so}(3)$ to $\mathbb{R}^3$ [42]. Fig. 4 depicts the simulated AC$_S$, neglecting the influence of the sidewalls. In this figure, the axes labels are according to the coordinate system shown in Figure 3. After applying a small curvature (3 m$^{-1}$) to the lower beam (illustrated in blue), due to the clamped ends, the upper beam (depicted in red) undergoes deformation, as shown in Fig. 4b. Distance between the beams after applying the curvature is shown in Fig. 4c. From this figure, it can be observed that in certain regions, the distance between the beams notably decreases, approaching zero, after experiencing small curvature.

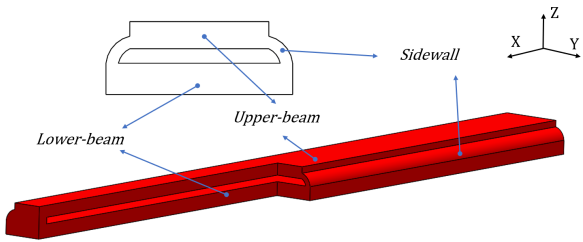

**Fig. 3.** The proposed acoustic $AC_S$ structure.

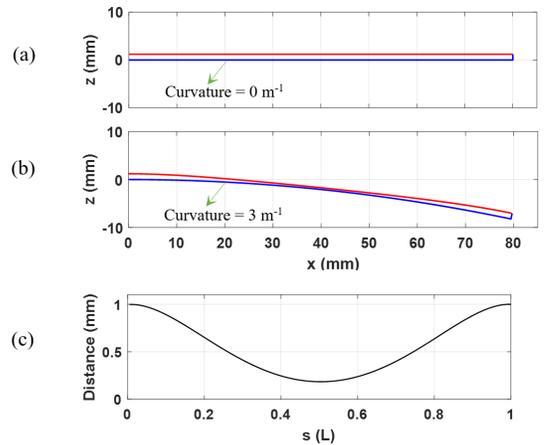

(a)

(b)

(c)

**Fig. 4.** The simulated channel (a) before and (b) after applying the curvature, with the (c) distance between the beams.

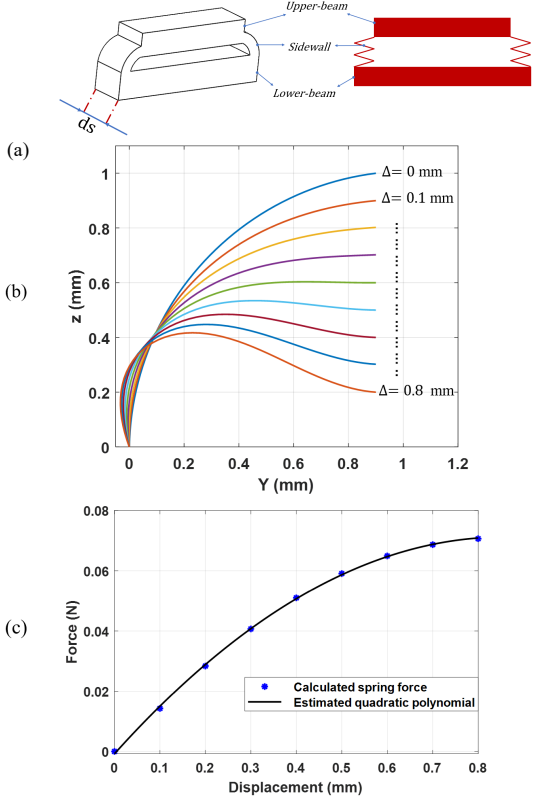

**Fig. 5.** (a) A longitudinal segment of the AC$_S$ and its schematic diagram, and (b) simulation results for the sidewall of the segment, with the (c) relation between the displacement of the sidewall tip and the resulting spring force.

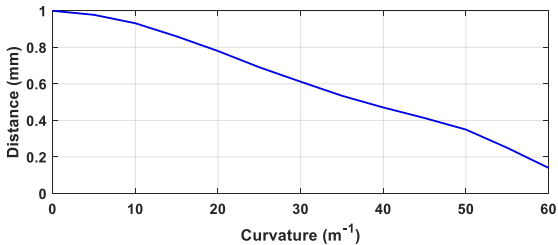

**Fig. 6.** The minimum channel height in different curvature.

The sidewalls added to the sensor regulate the distance between the beams during bending, enabling the sensor to effectively measure both large and small curvatures. At each longitudinal segment of the AC$_S$ (Fig. 5a), the sidewall functions as a spring, exerting a resisting force ($\boldsymbol{\rho}$) against the beams as they approach each other due to the imposed curvature. The simulation result for the sidewall of the segment is shown in Fig. 5b. When the upper beam moves toward the lower beam, it causes a displacement $\Delta$ at the tip point of the sidewall, and this segment reacts by applying an upward spring force to the upper beam. The relationship between displacement and the spring force is depicted in Fig. 5c. This force can be approximated as a quadratic polynomial function of displacement (shown in black), which can then be utilized in the simulation of the entire channel.

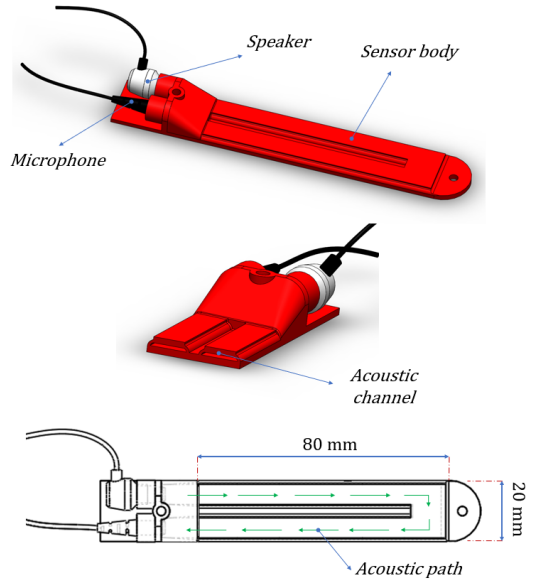

**Fig. 7.** The design of the soft acoustic curvature sensor.

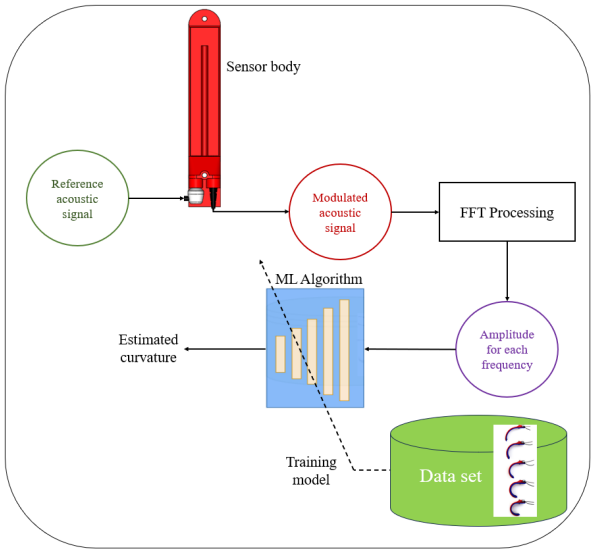

**Fig. 8.** The schematic of the proposed curvature sensing method.

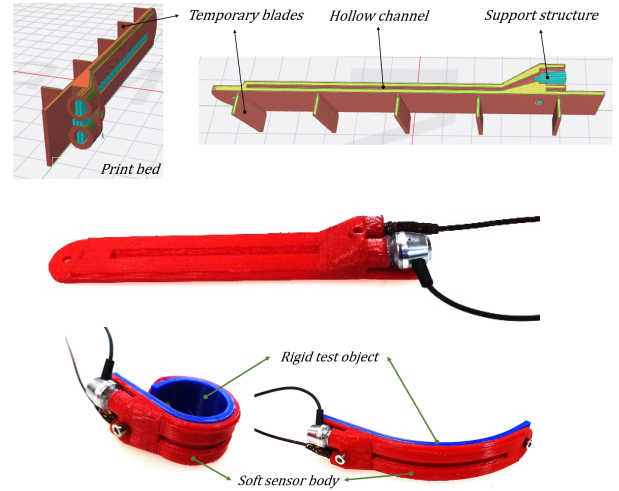

**Fig. 9.** (a) The build orientation in the printing process, (b) The fabricated sensor, and (c) the test objects to apply various curvatures to the sensor body.

The simulation results for the minimum channel heights after applying various curvatures are presented in Fig. 6. Therefore, based on the results, it is anticipated that the proposed channel structure will lead to a consistent decrease in the height of the channel in certain areas with increased curvature, effectively modulating the sound signals.

Our proposed SAC sensor design is depicted in Fig. 7. It features a flat shape with a small thickness to make it flexible (not stretchable) and readily adaptable for various applications. To extend the curvature range and enhance usability, the sensor body incorporates a U-turn in the acoustic channel, allowing both the microphone and speaker to be positioned on the same end. This configuration permits greater bending within the same channel length, enabling the sensor to accommodate a wider range of curvatures.

### C. Curvature sensing

The proposed sensor operates by having the speaker continuously emit acoustic waves at specific frequencies through the channel, while the microphone captures these waves. Changes in the curvature of the sensor result in alterations in bending and the height of the ACs, consequently affecting the amplitude of acoustic waves. To detect curvature from the microphone recordings, a mapping between the channel deformations caused by a change in curvature and the resulting sound modulations needs to be established. However, deriving an analytical model for this mapping is overly complex, especially for deformable soft objects [29]-[32].

We employed ML methods to establish the relationship between sensor curvature and the resulting sound amplitude modulations, as this relationship is highly nonlinear and complex to model analytically. To train the ML model, we created a dataset by wrapping the sensor body around circular test objects of known radii. For each curvature, the speaker emitted sound waves at different chosen frequencies, and the microphone recorded the corresponding samples. The FFT of these recorded samples was computed and processed to create a dataset for training various ML models. Model selection was guided by the validation error, with Gaussian Process Regression (GPR) yielding the lowest validation error (see Table I) and thus being chosen as the most suitable model. We then utilized GPR to estimate the curvatures of unknown objects based on the microphone recordings. The overall methodology is illustrated in Fig. 8.

(b)

(a)

## III. Experimental procedure

The soft sensor body is fabricated using 3D printing with flexible Ninjaflex filament on a RAISE3D Pro2 printer. To avoid the need for support structures inside the acoustic channel during printing, the sensor body is printed in an on-edge orientation, as shown in Fig. 9a. Temporary stabilizing blades are incorporated into the design to prevent deformation during printing. These blades are connected to the lower beam by their sharp ends and can be easily removed after printing. Additionally, minor support structures at the beginning of the channel (shown in green) are used easily removed post-printing. Fig. 9b shows the fabricated sensor body. The cost of the fabricated sensor was below 40 $.

Sine waves ranging from 200 to 2000 Hz were chosen as the reference acoustic signal due to their sensitivity to the changes in channel height. The reference signal was generated using Audacity software. To apply various curvatures to the sensor and generate the training dataset, test objects with known curvatures, printed from PLA, were utilised, as depicted in Fig. 9c. After attaching the sensor to the test objects, a continuous reference acoustic signal was played through the speakers, while the microphone recorded 50 samples of the received sound signal. The FFT algorithm is used to determine the amplitudes of the acoustic signals recorded by the microphone. The resulting dataset comprises curvature values and the amplitudes of the selected frequencies. 13 curvatures ranging from 0 to 60 $m^{-1}$, in steps of 5 $m^{-1}$, were applied to the sensor (Fig. 10a), resulting in a dataset consisting of 6500 samples. Fig. 10b illustrates the variation in amplitudes with curvature for each frequency.

Some frequencies exhibit an increase in amplitude within the range of 0 to 25 $m^{-1}$, as the decrease in the channel's height is negligible in this range of curvatures and the effect of channel's bending is more pronounced. As curvature increases, the impact on channel's height becomes more significant, leading to a decrease in amplitude for all frequencies. The diversity of amplitudes across different frequencies enhances the model training performance and validates the utility of FFT data for predicting curvature values. Various tests demonstrated that maximum variation of the recorded amplitudes for each frequency at each curvature was below 3%. Subsequently, various regression models were trained using the datasets with a partition of 90:10 and 10-fold cross-validation. We observe RQ-GPR is the model with the lowest validation error chosen for predicting the curvature using our sensor.

## IV. Results and Discussion

Table I summarizes the validation Root Mean Squared (RMS) error for each trained model. The results show that the Rational Quadratic Gaussian Process Regression (RQ-GPR) model consistently delivered the most reliable performance for the SAC sensors. To assess the accuracy of the proposed curvature measurement method, we subjected the sensor to validation tests using rigid test objects of known curvatures. Importantly, a unique set of test objects, distinct from those used during the training process, was employed for validation. Each test was performed five times for each test object and each sensor.

### TABLE I
### Comparison of the regression models

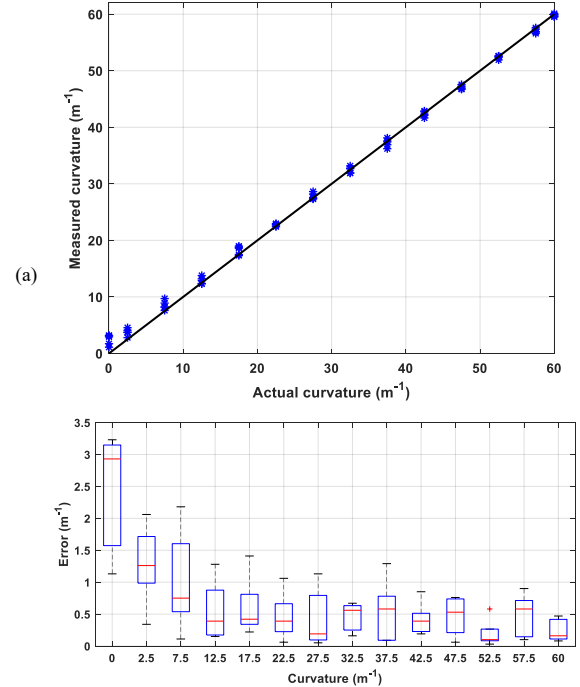

**Fig. 11.** (a) Comparison of the actual and measured curvatures and (b) the error deviations in curvature measurement.

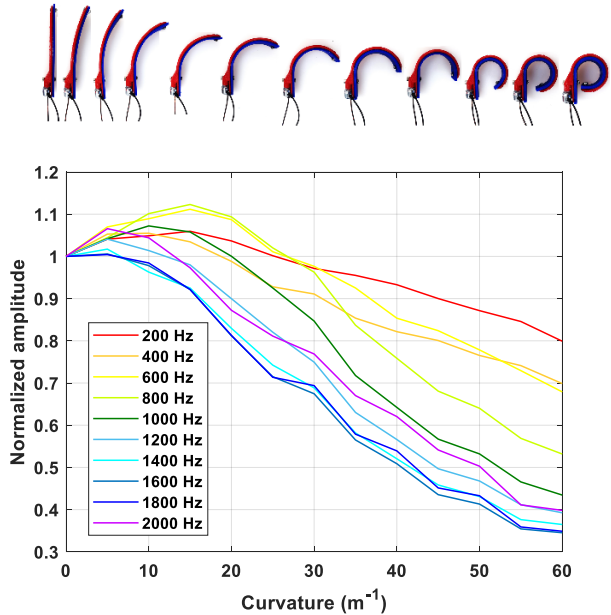

**Fig. 10.** (a) Curvature test objects ranging from 0 to 60 $m^{-1}$. (b) Variation of the recorded amplitude with curvature for each frequency

| | Medium Gaussian SVM | 1.19 |
| | Coarse Gaussian SVM | 1.48 |
| Gaussian Process Regression (GPR) | Squared Exponential GPR | 0.44 |
| | Matern 5/2 GPR | 0.26 |
| | Exponential GPR | 0.25 |
| | **Rational Quadratic GPR** | **0.22** |

In Fig. 11a, experimental observations are shown as scattered points, with the ideal 1:1 line depicted as a solid line. Corresponding errors are presented in Fig. 11b. The results indicate that while the sensor's errors are relatively large (∼<3 m$^{-1}$) at smaller curvatures (curvature < 10 m$^{-1}$), the sensor performs well for curvatures exceeding 10 m$^{-1}$, with errors reduced to ∼<1 m$^{-1}$. Table II presents a summary of the minimum, average, and maximum errors for each sensor at various curvatures. From the results, the maximum curvature estimation error using the proposed method consistently stays bounded < 3.5 m$^{-1}$. For curvatures exceeding 10 m$^{-1}$, the error consistently drops below 1.5 m$^{-1}$.

TABLE II
MEASUREMENT RESULTS

| Curvature (m$^{-1}$) | Measurement error (m$^{-1}$) | | | Curvature (m$^{-1}$) | Measurement error (m$^{-1}$) | | |
|---|---|---|---|---|---|---|---|
| | Min | Avr | Max | | Min | Avr | Max |
| 0 | 1.13 | 2.42 | 3.23 | 32.5 | 0.16 | 0.45 | 0.67 |
| 2.5 | 0.34 | 1.29 | 2.06 | 37.5 | 0.09 | 0.53 | 1.29 |
| 7.5 | 0.11 | 1.02 | 2.18 | 42.5 | 0.19 | 0.41 | 0.85 |
| 12.5 | 0.15 | 0.54 | 1.28 | 47.5 | 0.06 | 0.46 | 0.76 |
| 17.5 | 0.22 | 0.60 | 1.41 | 52.5 | 0.03 | 0.19 | 0.58 |
| 22.5 | 0.06 | 0.46 | 1.06 | 57.5 | 0.12 | 0.47 | 0.90 |
| 27.5 | 0.05 | 0.43 | 1.13 | 60 | 0.08 | 0.24 | 0.47 |

In the next experiment, the SAC sensor is integrated with the back-beam of a Fin Ray Effect (FRE) finger [43] to provide feedback from its shape (Figure 12a). The curvature measured by SAC sensor is compared with those measured by our perception curvature estimation [44] that uses a Direct Linear Transformation approach based on two cameras (DSC-RX100M4, Sony) in stereo configuration. The gripper grasped several objects of various shapes and the comparison of the curvature measurements after gripping objects is given in Figure 12b. The results show the error is below 3.5 m$^{-1}$ with larger errors occurring in smaller curvatures.

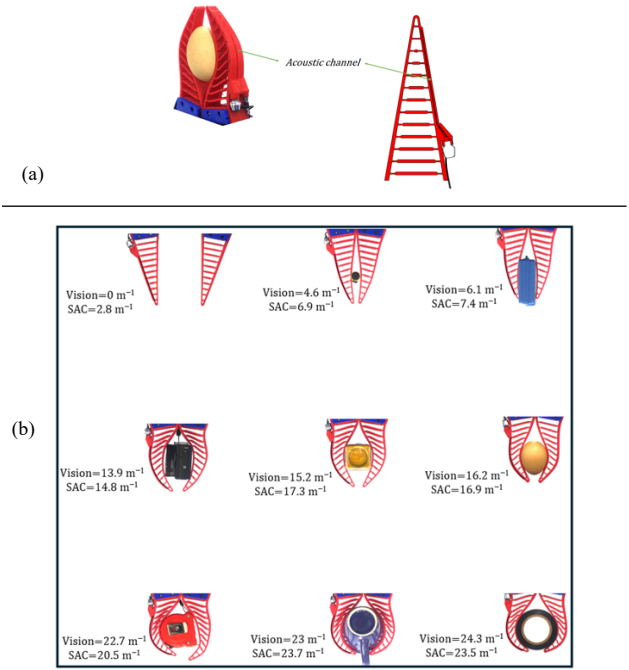

(a)

(b)

**Fig. 12.** (a) The FRE gripper with the embedded acoustic channel. (b) Comparison of the measured curvatures using vision system and the proposed SAC sensor.

Expanding the length of the acoustic channels is anticipated to enhance measurement accuracy by introducing more modulation effects on the acoustic waves. Consequently, sensors with longer sensing bodies are expected to exhibit superior performance, which will be a focus of future research. Additionally, implementing a more intricate channel path with multiple U-turns may further improve accuracy.

It is important to note that applying external forces to the acoustic channels, can lead to measurement errors due to channel deformation. Experimental investigations showed that applying a normal force of 1N (distributed over area of 1cm$^2$) can increase the average measurement errors by up to 18%. Therefore, it is crucial to ensure that the sensor body remains isolated from external forces, such as avoiding placing it on the contact surface of soft grippers. The sensor, when properly sealed, is robust against environmental acoustic noises; however, loud acoustic noises can affect its performance. Experimental tests showed that with white noise below 60 decibels, the changes in measurement error are negligible (below 4%), but noise at 120 decibels increases the average measurement error by up to 32%. Additionally, twisting the sensor body from strait condition (curvature = 0 m$^{-1}$) by 30 degrees, showed an average error increase of 12%.

While our proposed design works for 1-D curvature and the other dimension should be planner, the measurement of non-planar surfaces can be explored in future research. Future works also include: (1) exploring the effects of varying sensor body sizes and incorporating complex acoustic channels with multiple U-turns to assess their impact on sensor performance; (2) exploring different material for the sensor body; (3) integrating the SAC sensor onto the bodies of soft robots or within their structures to provide real-time curvature feedback; and (4) using SAC sensor for handling soft fruits similar to AST sensor [45]. This advancement has the potential to facilitate the estimation of overall shapes using the PCC kinematic model, enhancing the capabilities of soft robotic systems. When designing a new sensor with a different shape or size, or when integrating it into new structures, the training procedure must be repeated.

## V. CONCLUSION

In this paper, we introduced the Soft Acoustic Curvature (SAC) sensor, a novel device designed to measure the curvature of soft bodies using integrated audio components and a novel acoustic channel within a flexible structure. The integration of machine learning models learns to effectively estimate curvature. Through comprehensive experimentation and analysis, we demonstrated that the SAC sensor effectively modulates sound waves in response to changes in curvature, with measurement errors remaining within 3.5 m$^{-1}$ for curvatures ranging from 0 to 60 m$^{-1}$. The SAC sensor offers several practical advantages, including low cost, flexibility, adaptability to various shapes and spaces, and minimal impact on the stiffness of the attached soft body. These features make it highly suitable for applications in soft robotics, such as shape measurement for continuum manipulators, soft grippers, and wearable devices.